\documentclass{appolb}
\usepackage{graphicx}
 \usepackage{amsmath}
 \usepackage{amssymb}
 \usepackage{bbold}
 \usepackage{bm}
 \usepackage{slashed}
\usepackage{float}
\usepackage[dvipsnames]{xcolor}
 \usepackage{textcomp}
\begin{document}
% \eqsec  % uncomment this line to get equations numbered by (sec.num)
\title{Quark mass function from a OGE-type interaction in Minkowski space%
\thanks{Presented at Excited QCD 2019, Schladming, Austria.}%
% you can use '\\' to break lines
}
\author{Elmar P. Biernat$^{1}$, Franz Gross$^{2}$, M. T. Pe\~na$^{1}$, Alfred Stadler$^{1,3}$, Sofia Leit\~ao$^{1}$
\address{
$^{1}$CFTP and Departamento de F\'isica, Instituto Superior T\'ecnico, Universidade de Lisboa, Avenida Rovisco Pais 1, 
1049 Lisboa, Portugal
% \email{elmar.biernat@tecnico.ulisboa.pt}
\\
$^{2}$Thomas Jefferson National Accelerator Facility (JLab), Newport News, VA 23606, USA and College of William and Mary, Williamsburg, Virginia 23188,
USA
\\
$^{3}$Departamento de F\'isica, Universidade de \'Evora, 7000-671 \'Evora, Portugal}}
% \author{First Author, second Author
% \address{affiliation}
% \\
% {Third Author of different affiliation
% }
% the Name(s) of other Author(s)
% \address{affiliation}
% }
\maketitle
\begin{abstract}
We present results for the quark mass function in Minkowski space calculated from an interaction kernel that consists of an effective one-gluon-exchange and a constant interaction. We analyze the gauge dependence of our results and compare them in the spacelike region to the available lattice QCD data.
\end{abstract}
  
\section{Introduction}

QCD at low energies requires an essentially non-perturbative treatment which makes the theoretical description of hadrons difficult. For the light mesons, in particular the pion, the implementation of dynamical chiral symmetry breaking (D$\chi$SB) is indispensable. In this work we present recent results on the quark self-energy obtained in the ``Covariant Spectator Theory'' (CST)~\cite{Gross:1969eb,Gross:1982}, which has already been applied successfully to mesons previously~\cite{Gross:1991te,Gross:1991pk,GMilana:1994,Savkli:2001os,PhysRevD.98.114033,Biernat:2014jt,pionff:2014,pionff:2015,Biernat:2014xaa,Biernat:2012ig,Leitao:2017mlx,Leitao:2017it,Leitao2017,Leitao:2014}. The CST is a covariant approach formulated in Minkowski space and related to the Bethe-Salpeter/Dyson Schwinger formalism~(for a recent review, see~\cite{Eichmann:2016bf}). It uses a quark-quark interaction kernel that includes, in addition to the one-gluon-exchange (OGE) kernel, a covariant phenomenological generalization of a linear confining potential.

\section{Quark self-energy in CST}
The dynamical quark mass generation is described by the Dyson equation for the dressed quark propagator given by
\begin{equation}
S(p)=S_0(p)- S_0 (p) Z_2 \Sigma({\slashed p}) S(p)\,,
\label{eq:DE}
\end{equation}  
where { $S_0(p)=\left(m_0-\slashed{p}- \mathrm i\epsilon\right)^{-1}$} and $S(p)=\left(m_0-\slashed{p}+Z_2 \Sigma(\slashed{p})- \mathrm i\epsilon\right)^{-1}$ are the bare and dressed quark propagators, respectively, with $m_0$ the bare (current) quark mass, $Z_2$ a renormalization constant, and $\Sigma(\slashed{p})$ the quark self-energy which can be written in terms of invariant functions as 
\begin{equation}\Sigma(\slashed{p})=A(p^2)+\slashed{p}\,B(p^2)\, . \label{eq:se}
\end{equation} The quark mass function and wave function normalization are related to the self-energy by
\begin{eqnarray}
M(p^2)=Z(p^2) \left[ m_0+Z_2A(p^2) \right] \quad {\rm and}\quad Z(p^2)=\frac{1}{1-Z_2B(p^2)} 
\, , \label{eq:ZM}
\end{eqnarray}
respectively. 
One of the central assumptions of the CST is the existence of a real mass pole of the dressed quark propagator at $p^2=m^2$, identified as the constituent quark mass $m$, such that $M(p^2)$ and $Z(p^2)$ satisfy
\begin{eqnarray}
M(m^2)=m
\quad {\rm and}\quad Z(m^2)=Z_2\left[1-2m \frac{\mathrm d M(p^2)}{\mathrm d p^2}\bigg|_{p^2=m^2}\right] \,,\label{eq:oscondition} 
\end{eqnarray} 
respectively. In CST, the zero-components of loop momenta are integrated by calculating only the residues of the quark propagator poles, such that the CST self-energy (times $Z_2$) is given by
\begin{eqnarray}
Z_2\Sigma (\slashed p)= \frac {Z_2^2}{ 2}\sum_{\sigma=\pm}\int_{\bf k} \mathcal V (p,{ \hat k_\sigma })  \left(\frac{m+\hat {\slashed{k}}_\sigma}{2m}\right)\,,\label{eq:DEk01}
\end{eqnarray}
where $\int_{\bf k} \equiv  \int  \frac{\mathrm d^3 {\bf k}}{(2\pi)^3} \frac{m}{E_k}$, $\sigma=\pm$ labels the positive- and negative-energy on-shell momenta (corresponding to the positions of the quark propagator poles), $\hat k_\sigma=(\sigma E_k,{\bf k})$, with  \mbox{$E_k=\sqrt{m^2+{\bf k}^2}$} and $\mathcal V (p,{ \hat k_\sigma })$ is the interaction kernel given by
 \begin{eqnarray}
 {\cal V}(p,\hat k_\sigma) &=&\left(\frac14\sum_a\lambda_a\otimes \lambda_a\right)
 \left\{\left[\left(\mathbf{1}\otimes \mathbf{1}+\gamma^5\otimes\gamma^5\right)\right]V_{\ell} (p,\hat k_\sigma)\right.\nonumber\\& &-\left.\gamma_\mu\otimes\gamma_\nu \left[\Delta_{\mathrm g}^{\mu\nu}(q_\sigma^2)V_{\mathrm g}(p,\hat k_\sigma)+\Delta_{\mathrm c}^{\mu\nu}(q_\sigma^2)V_{\mathrm c}(p,\hat k_\sigma)\right] \right\}\,.\label{eq:Vdecomp}
 \end{eqnarray}
 Here $q_\sigma= p-\hat k_\sigma$, ${V}_{\ell}$ is a covariant generalization of a linear
confining potential, 
\begin{eqnarray}
 {V}_{\mathrm g}(p,\hat k_\sigma)&=&-4\pi\alpha_{\rm s} \frac{ g(y)}{(p-\hat k_\sigma)^2}
 \label{eq:OGEgauge0}
\end{eqnarray} is the OGE interaction with $\alpha_{\rm s}$ the \emph{unrenormalized} strong coupling constant, $g(y)$ a regularization form factor depending on the covariant variable $y^2=\frac{(p\cdot k)^2}{p^2k^2}$,
\begin{eqnarray}
{V}_{\rm c}(p, \hat k_\sigma)&=&\frac{C E_k}{2m}(2\pi)^3\delta^3\Big({\bf k}-\frac{m}{\sqrt{p^2}}\, {\bf p}\Big) % 
h(p^2)
 \, , \label{eq:Ckernel}
\end{eqnarray}
is a covariant form of a constant potential with $C$ the unrenormalized strength and $h$ is a strong quark form factor normalized as $h(m^2)=1$. The $\Delta^{\mu\nu}$'s in Eq.~(\ref{eq:Vdecomp}) are factors given in general linear covariant gauge specified by the gauge parameter $\xi$ and chosen in this work as
\begin{eqnarray}
 \Delta_{\mathrm c}^{\mu\nu}(q_\sigma^2)= \mathrm g^{\mu\nu}-(1-\xi)\frac{q_\sigma^\mu q_\sigma^\nu}{q_\sigma^2}\quad \text{and} \quad  \Delta_{\mathrm g}^{\mu\nu}(q_\sigma^2)=  -\frac{q_\sigma^2}{M_{\rm g}^2+|q_\sigma^2|}\Delta_{\mathrm c}^{\mu\nu}(q_\sigma^2)\,.\label{eq:ansaetzedressing}
\end{eqnarray}
Notice that this choice effectively gives the gluon a finite mass $M_{\rm g}$ and replaces $q^2\to -|q^2|$, which removes the singularity in the gluon propagator. Further recall~\cite{Biernat:2014xaa} that the linear confining part of ${\cal V}$ gives no contribution to the CST self-energy (\ref{eq:DEk01}), which simplifies the calculation substantially.
 \section{Results and Discussion}
In the quark's rest frame, where $p=\{p_0,{\bf 0}\}$, the OGE contributions to the invariant self-energy functions are 
\begin{eqnarray}
Z_2 A_{\rm g}(p_0^2)&=&\frac{8\pi \alpha_{\rm s}^{\rm r} }{3}\,m\sum_\sigma\int_{\bf k}
\frac{(3+\xi)g(y) }{M_{\rm g}^2+|q_\sigma^2|}
\, ,
\nonumber\\
Z_2 B_{\rm g}(p_0^2)&=&- \frac{8\pi\alpha_{\rm s}^{\rm r }}{3}  \sum_\sigma\int_{\bf k}\frac{g(y)}{M_{\rm g}^2+|q_\sigma^2|}\left\{\frac{\sigma (3-\xi)E_k}{p_0}\,+ \frac{2(1-\xi){\bf k}^2}{ q_\sigma^2}\right\}\,
 \label{eq:OGEcontr}
\end{eqnarray}
where $\alpha_{\rm s}^{\rm r }=Z_2^2\alpha_{\rm s}$ is the \emph{renormalized} strong coupling constant. The renormalization of $\alpha_{\rm s}$ arises from a factor of $\sqrt{Z_2}$ attached to each quark line either entering or leaving an interaction vertex. For $M_{\rm g}$ we take the value $M_{\rm g}=0.6$ GeV and for $g(y)$ we chose the form
\begin{eqnarray}
g(y)&=&
\frac{\Lambda_{\rm g}^{8}}{\Lambda_{\rm g}^{8}+m^8(y^2-1)^4} \, ,
 \label{eq:hg}
\end{eqnarray}
where $\Lambda_{\rm g}$ is an adjustable scale parameter. It can be shown analytically that the on-shell equations~(\ref{eq:oscondition}) for the OGE contributions (\ref{eq:OGEcontr}) are \emph{independent} of $\xi$, and so are the constituent quark mass $m$ and $Z_2$. When Eq.~(\ref{eq:oscondition}) is solved in the chiral limit ($m_0=0$) with $m=0.3$ GeV and $\Lambda_{\rm g}\lesssim 2$ GeV for the OGE kernel alone, then the value for $\alpha_{\rm s}^{\rm r}$ turns out to be unnaturally large as compared to the approximate value known from experiment. The additional constant kernel, which can be regarded as a correction to the OGE contribution, solves this issue and leads to realistic values of our model parameters. Its contributions are
\begin{eqnarray}
Z_2A_{\rm c}(p^2)&=&\frac14(3+\xi) {C}^{\rm r} \,h(p^2) \, ,\quad
Z_2 B_{\rm c}(p^2)=0\,  ,\label{eq:AcBc}
\end{eqnarray}
where $C^{\rm r }=Z_2^2C$ is the \emph{renormalized} strength. If $Z_2A_{\rm c}(p^2)$ is to satisfy~(\ref{eq:oscondition}) in the chiral limit in arbitrary gauge, then ${C}^{\rm r}\to 4m/(3+\xi)$, and if the constant kernel is to supplement the OGE kernel, then it is appropriate to choose 
$h(p^2)=A_{\rm g}(p^2)/A_{\rm g}(m^2)$.

\begin{figure}[H]
\centerline{%
\includegraphics[width=9cm]{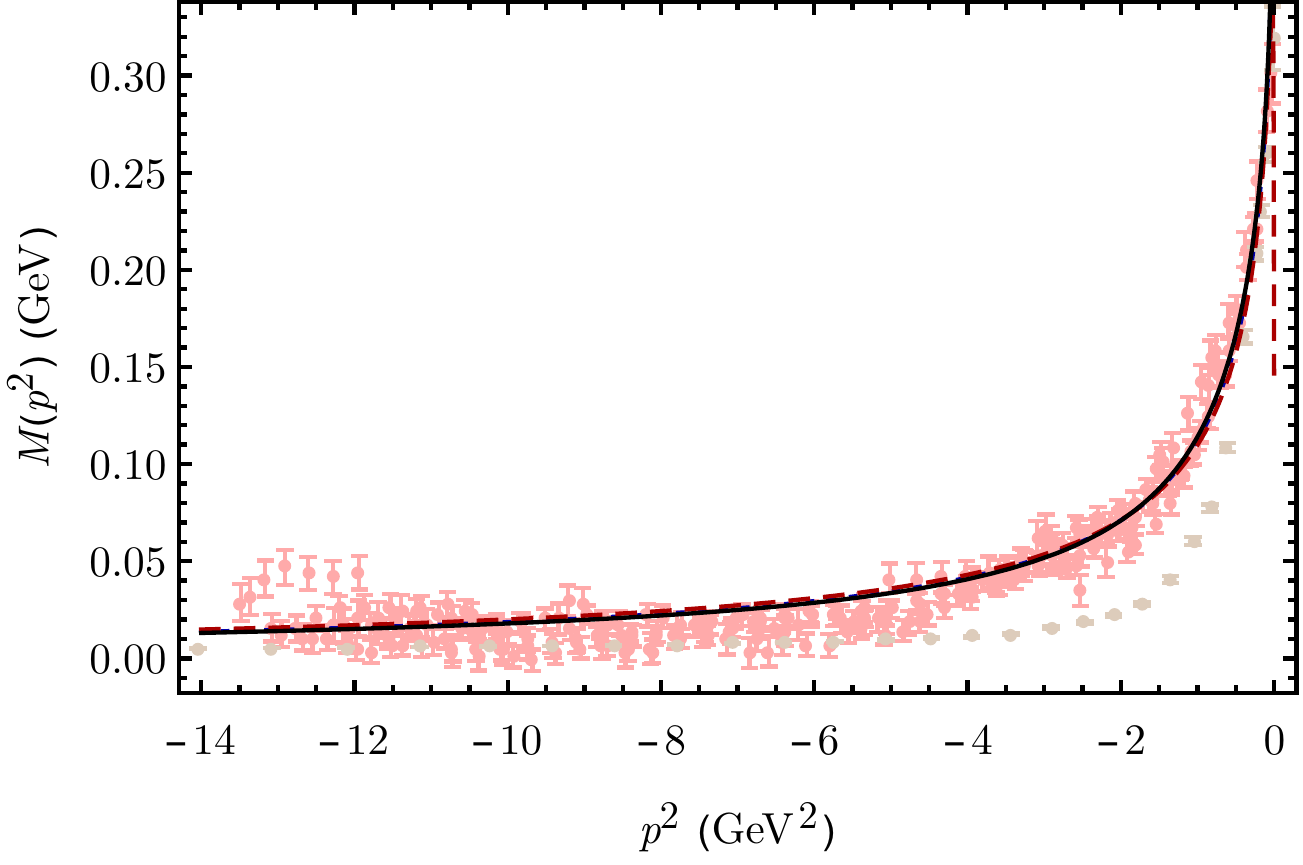}}
\centerline{%
\includegraphics[width=9cm]{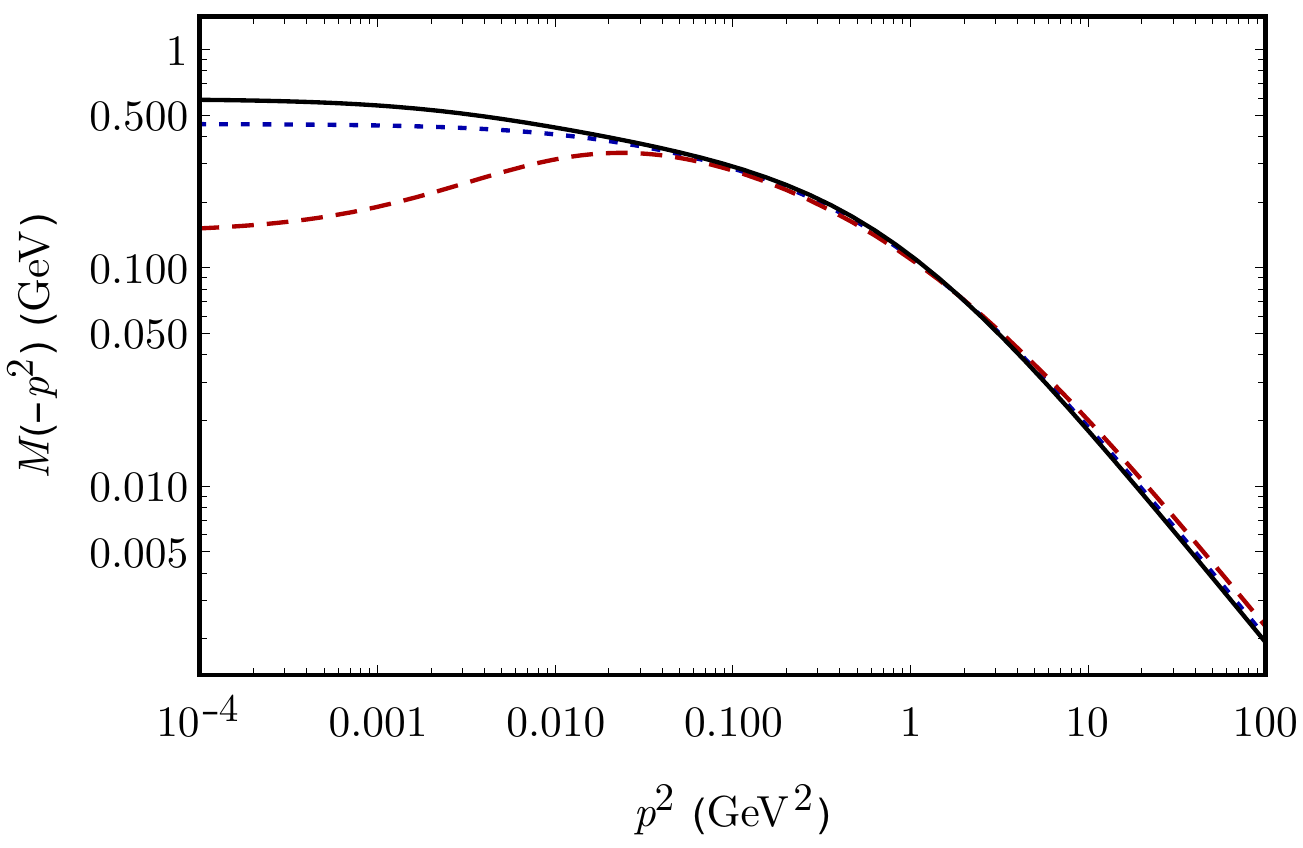}}
\centerline{\includegraphics[width=9cm]{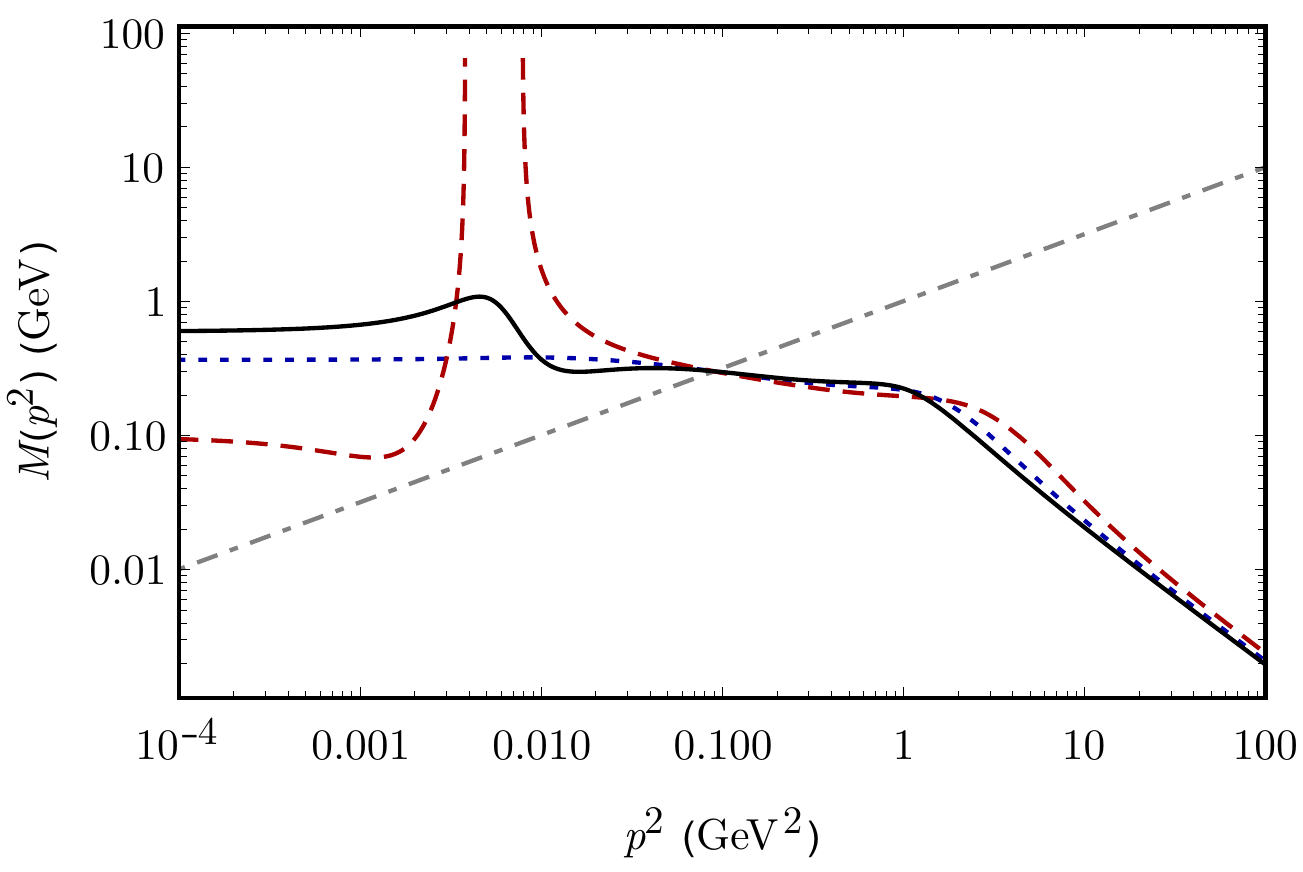}}
\caption{The mass function (in GeV) vs. $p^2$ (in GeV$^2$) for $\xi = 0$ and $\Lambda_{\rm g}=0.9$ GeV (red dashed line), $\xi=1$ and $\Lambda_{\rm g}=0.6$ GeV (blue dotted line), and $\xi = 3$ and $\Lambda_{\rm g}=0.45$ GeV (black solid line). The top panel shows the spacelike region with lattice QCD data taken from~\cite{Bowman:2005vx} (red data points) and~\cite{Oliveira:2018lln} (brown data points). The middle and bottom panel show the spacelike and timelike regions on logarithmic scales, respectively, where the intersection of the curves with $\sqrt{p^2}$ (gray dotdashed line) marks the $\xi$-independent on-shell point $p^2=m^2$.}
\label{fig:MC&OGE}
\end{figure}

Because each contribution~(\ref{eq:OGEcontr}) and~(\ref{eq:AcBc}) satisfies~(\ref{eq:oscondition}) separately, this suggests to write the total result as a linear combination of these contributions:
\begin{eqnarray}
Z_2 A(p^2)= \left[\eta+\frac{m(1- \eta)}{Z_2 A_{\rm g}(m^2)} \right] Z_2  A_{\rm g}(p^2)\,,\quad Z_2 B(p^2)= \eta Z_2  B_{\rm g}(p^2)\,,
 \label{eq:totalcontr}
\end{eqnarray}
where $\eta$ is a mixing parameter chosen to maintain the effective OGE strength $\eta\alpha_{\rm s}^{\rm r}=0.5$ regardless of the choice of $\Lambda_{\rm g}$, which is roughly adjusted to agree with the lattice QCD data~\cite{Bowman:2005vx}. The results for the chiral-limit quark mass function are summarized in Fig.~\ref{fig:MC&OGE}.

Notice that in Landau gauge ($\xi = 0$) there is a pole in the mass function at some small timelike $p^2$ where $Z_2B(p^2)=1$. Such poles are not present in the Feynman ($\xi = 1$) and Yennie ($\xi = 3$) gauge results. Further notice that, in order to make the dressed gluon propagator non-singular we have introduced particular gluon dressing functions in Eq.~(\ref{eq:ansaetzedressing}). The disadvantage of this choice is that the mass function develops a discontinuity at $p^2=0$ (not displayed in the figure). Fortunately, the size of the discontinuity depends on the gauge and we find that the quark mass function is continuous at $p^2=0$ only for the Yennie gauge, which therefore constitutes the preferred gauge in this calculation. 
In the timelike region there is a strong dependence of the quark mass function results on the gauge, however on-mass-shell results are gauge independent -- a natural feature of the CST. In the spacelike region the gauge dependence is very weak, allowing us to use the existing lattice QCD data for the mass function to calibrate the model. We now have a mass function that can be used in meson calculations where the quark dressing and the quark-quark interaction are completely consistent.

\subsection*{Acknowledgments}
This work was funded in part by Funda\c c\~ao para a Ci\^encia e a 
Tecnologia (FCT) under Grants No. CFTP-FCT (UID/FIS/00777/2013), No. SFRH/BPD/100578/2014, and No. SFRH/BD/92637/2013. F.G. was supported by the U.S. Department of Energy, Office of Science, Office of Nuclear Physics under contract DE-AC05-06OR23177.

%
% \section{Subsection}
% The text...

%uncomment the following lines to place a figure
%\begin{figure}[htb]
%\centerline{%
%\includegraphics[width=12.5cm]{Fig1}}
%\caption{Plot of ...}
%\label{Fig:F2H}
%\end{figure}

%%%%%%%%%%%%%%%%%%%%%% BIBLIOGRAPHY %%%%%%%%%%%%%%%%%%%%%%%%

%   
%   \bibliographystyle{h-physrev3}
%   \bibliography{PapersDB-v2-3}

\end{document}